# Elastic Free Energy Drives the Shape of Prevascular Solid Tumors

K. L. Mills[1]*, Ralf Kemkemer[1,2], Shiva Rudraraju[3], and Krishna Garikipati[3,4]*

*Affiliations:*
[1]Max Planck Institute for Intelligent Systems, Department of New Materials and Biosystems, Heisenbergstrasse 3, 70569 Stuttgart, Germany.
[2]Reutlingen University, Alteburgstrasse 150, 72762 Reutlingen, Germany.
[3]University of Michigan, Department of Mechanical Engineering, 2350 Hayward, Ann Arbor, MI 48109, USA.
[4]University of Michigan, Department of Mathematics, 530 Church Street, Ann Arbor, MI 48109, USA.

*Correspondence:
K. L. Mills
Max Planck Institute for Intelligent Systems
Department of New Materials and Biosystems
Heisenbergstrasse 3
70569 Stuttgart
Germany
+49 711 6893632
mills@is.mpg.de

Krishna Garikipati
Department of Mechanical Engineering
Deparment of Mathematics
2350 Hayward
Ann Arbor, MI 48109
USA
(734) 936-0414
krishna@umich.edu

*Keywords:* elasticity, cancer, tumor growth, finite element

**ABSTRACT**

It is well established that the mechanical environment influences cell functions in health and disease. Here, we address how the mechanical environment influences tumor growth, in particular, the shape of solid tumors. In an *in vitro* tumor model, which isolates mechanical interactions between tumor cells and a hydrogel, we find that tumors grow as ellipsoids, resembling the same, oft-reported observation of *in vivo* tumors. Specifically, an oblate ellipsoidal tumor shape robustly occurs when the tumors grow in hydrogels that are stiffer than the tumors, but when they grow in more compliant hydrogels they remain closer to spherical in shape. Using large scale, nonlinear elasticity computations we show that the oblate ellipsoidal shape minimizes the elastic free energy of the tumor-hydrogel system. Having eliminated a number of other candidate explanations, we hypothesize that minimization of the elastic free energy is the reason for predominance of the experimentally observed ellipsoidal shape. This result may hold significance for explaining the shape progression of early solid tumors *in vivo* and is an important step in understanding the processes underlying solid tumor growth.

# INTRODUCTION

Tumorigenesis and solid tumor growth are associated with altered mechanics in the tumor's environment such as increased matrix stiffness and growth-induced mechanical stress [1-4]. Such altered mechanical properties and stresses influence cancer cell behaviors such as growth, survival, organization [1-3], and tensional homeostasis [4], aid in invasion [5] and effect gene expression [6]. Additionally, the tumor's size and morphology may be affected [7-9].

For example, both *in vivo* and *in vitro*, solid tumors are often described as being ellipsoidal in shape. *In vivo*, the growth of solid tumors in ellipsoidal shapes is common across tumor classifications and tissue of origin [10-16]. This ellipsoidal growth of tumors can be emulated *in vitro*. In a model of the early, prevascular stage of tumor growth, tumor cells are embedded and allowed to grow in a tissue-mechanics-mimicking hydrogel [7-9]. In such an environment, tumor shape has been shown to suffer a loss of symmetry from spherical to ellipsoidal for two different tumor cell lines [7,9]. Since the hydrogel is biochemically inert to cellular attachment, the effect of the mechanical environment on tumor growth is isolated.

The widespread occurrence of the ellipsoidal tumor shape *in vivo* and the fact that it can be reproduced with multiple cell lines *in vitro* with only mechanical, not biochemical, constraints suggest that the mechanical environment strongly influences tumor shape development. However, the physical conditions driving ellipsoidal tumor growth have not yet been described. Here, we report on our study of the mechanics of tumor growth using an *in vitro* model of prevascular tumor growth in chemically inert hydrogels.

# MATERIALS AND METHODS

## Cell culture and (immuno)fluorescence stainings

Reagents were from Gibco (Carlsbad) unless otherwise stated. Cells—human colon adenocarcinoma (LS174T from ECACC, Porton Down)—were propagated in Bio-Whittaker Eagle's minimum essential medium (Lonza, Walkersville) supplemented with 10% fetal bovine serum, 1% Pen/Strep, and 1% MEM non-essential amino acids. Cells were split prior to becoming confluent using 0.25% Trypsin-EDTA and counted with a Z2 Coulter Counter (Beckman Coulter). For fluorescence imaging, cell nuclei were stained with DAPI (Serva Electrophoresis, Heidelberg) or Hoechst (Invitrogen), Alexa Fluor 568 Phalloidin (Invitrogen) was used for immunofluorescent staining of the actin, and Alexa Fluor 594-conjugated E-Cadherin antibody (Cell Signaling Technology).

**Agarose hydrogel preparation, tumor embedding, and growth conditions**

Agarose powder (type VII, Sigma, St. Louis) was dissolved in heated deionized water. The liquid agarose had twice the agarose concentration (in wt./vol.%) than intended for the experiment. Diluting with warm cell culture medium at the time of mixing with the cell suspension established the final concentration as well as a nutritive environment for the cells.

Cell-laden gels were maintained in 6-well cell-culture inserts with 1-μm porous membrane bottom (BD-Falcon, Franklin Lakes). A cell-free layer of agarose liquid (0.5 mL) was first deposited on the bottom of the well insert and allowed to gel before adding the cell-containing second layer (3 – 4 mL). After the second layer of agarose had gelled, cell culture medium was added both on top of the gel and in the well containing the insert (5 – 7 mL in total).

Preproduced tumor spheroids [17,18], formed by the hanging drop method [19], were directly injected with a micropipette tip into the gel after it was added to the well insert but before it gelled.

**Tumor Imaging**

After the spheroids were embedded in the gels, their development was monitored for time periods from three weeks up to two months by acquiring phase contrast images (Axio Observer.Z1, Carl Zeiss AG, Jena), generally at 48-hour intervals. The projections of the tumors were imaged in this manner. Only tumors that were well away from any boundary of the cell culture well were imaged. Since the walls of the well are angled, there is optical distortion preventing imaging of tumors growing adjacent to the walls. Tumors located within approximately one tumor radius away from the top or bottom of the well tended to grow with their major axes aligned with the gel boundaries, but the experimental setup was not conducive to precise measurements of boundary-tumor distances. It is for this reason that only tumors well away from the boundaries were measured and considered here.

Confocal fluorescence imaging was performed using a Leica TCS SP5 X on the DM 6000 CFS upright microscope (Leica).

Light sheet fluorescence microscope imaging (for Figure 1) was performed on a Zeiss Lightsheet Z.1 microscope with the help of Dr. C. Schwindling at the Zeiss Microscopy Labs in Munich, Germany and (for Figure 2) on a custom-built light sheet fluorescence microscope in the lab of Prof. H. Schneckenburger with the assistance of S. Schickinger at the University of Aalen, Germany [20].

**Finite element computations of tumor growth as a finite strain, nonlinear elasticity problem**

The foundation of our computations lies in the definition of the deformation gradient tensor, $\boldsymbol{F} = \boldsymbol{1} + \partial \boldsymbol{u}/\partial \boldsymbol{X}$, where $\boldsymbol{u}$ is the displacement field vector, and $\boldsymbol{1}$ is the isotropic tensor. To model the kinematics of growth, we adopt the elasto-growth decomposition (see Refs [21-23] and [24]) $\boldsymbol{F} = \boldsymbol{F}^e \boldsymbol{F}^g$, where the growth tensor $\boldsymbol{F}^g = diag(\alpha_1, \alpha_2, \alpha_3)$. Here $\alpha_1$, $\alpha_2$, $\alpha_3$ are growth ratios in the Cartesian directions defined by the three axes of a tumor that, in general is ellipsoidal in shape. We model the tumor and gel as soft, isotropic, nearly incompressible materials, using a neo-Hookean strain energy density function:

$$\psi(\boldsymbol{C}^e) = \frac{1}{2}\lambda(det\boldsymbol{C}^e - \boldsymbol{1}) - \frac{1}{2}\left(\frac{\lambda}{2} + \mu\right)\ln det\boldsymbol{C}^e + \frac{1}{2}\mu(tr\boldsymbol{C}^e - 3)$$

where $\boldsymbol{C}^e = \frac{1}{2}(\boldsymbol{F}^{eT}\boldsymbol{F}^e - \boldsymbol{1})$ is the right elastic Cauchy-Green tensor, and $\lambda, \mu$ are elastic constants called Lamé parameters, which are related to the Young's modulus, $E$, and Poisson ratio, $\nu$ by

$$\lambda = \frac{\nu E}{(1+\nu)(1-2\nu)}, and\ \mu = \frac{E}{2(1+\nu)}.$$

We have carried out finite element computations of the nonlinear, finite strain elasticity problem of the growth of initially stress-free spherical and ellipsoidal tumors in a constraining gel with the nearly incompressible neo-Hookean strain energy density function, using established methods [21-23,25]. The neo-Hookean function is derived from statistical mechanical principles accounting for the underlying Gaussian network of polymer chains [26], which form the gel and the extracellular matrix in the tumor. The bulk to shear modulus ratios for tumor and gel were taken to be $\kappa_{tum}/\mu_{tum} = \kappa_{gel}/\mu_{gel} = 50$, to represent the near incompressibility of both. This corresponds to $\lambda_{tum}/\mu_{tum} = \lambda_{gel}/\mu_{gel} = 49$. Our computations provide the Cauchy stress tensor, $\sigma_{ij}$, and the elastic deformation gradient, $F^e_{ij}$. The latter enables us to compute the elastic free energy.

The initial tumor was modeled as an ellipsoid of semi-axes $a_1, a_2, a_3$, where $a_1, a_2$ lay in the range 50—72 $\mu$m, while $a_3$ was determined by requiring all the tumors to have a fixed initial volume, imposed by specifying $a_1 a_2 a_3 = 1.25 \times 10^5$ $\mu m^3$. The initial ellipsoids thus had aspect ratios of axes varying between 1 and 3. The ellipsoids were embedded in a cubically shaped gel of side 2000 $\mu$m. The tumor-gel interface allows transmission of normal and tangential tractions. This models the effect of bonding between the ECM deposited by the tumor cells, and the gel. The boundary value problem of nonlinear elasticity was driven by specifying the growth tensor to be isotropic, $\alpha_1 = \alpha_2 = \alpha_3$, with final growth volume ratio $\det \boldsymbol{F}^g = \alpha_1 \alpha_2 \alpha_3 = 5$ or $10$. We specified displacement boundary conditions, $\boldsymbol{u} \cdot \boldsymbol{n} = 0$ on five of the six boundaries of the gel, while on the sixth we used traction-free conditions $\boldsymbol{\sigma n} = \boldsymbol{0}$. Thus, we modeled the gel in a well with one free surface, as in our experiments. We confirmed that all stresses decayed by at least two orders of magnitude at the gel boundaries, relative to their maxima, thus ensuring that the extent of the gel and its shape had no influence on the computations.

In the finite element computations, the restriction of the tumor-gel interface to element edges would have two drawbacks: (a) It would require excessive mesh refinement to approximate the curved interface. (b) Even with a high degree of mesh refinement, stress singularities would arise at the edges and vertices of the hexahedral elements that would lie at the tumor-gel interface, because of the discontinuities in strain (from tumor growth) and elastic constants. Both these limitations can be mitigated if the finite element formulation can be extended to allow the tumor-gel interface to intersect an element. Such methods are well-known in the finite element literature. Our implementation is based on the enhanced strain formulation described in Garikipati & Rao (2001, [25]), which we have extended to three-dimensional hexahedral elements. We note that this method allows the tensors $\boldsymbol{F}$ and $\boldsymbol{F}^g$ to be discontinuous within elements that contain the tumor-gel interface, as dictated by the mathematically exact kinematics of the problem.

We have implemented the finite element formulation outlined above in the open source code deal.ii [27]. The stress fields reported in Figure 5 and the energy surface plots of Figure 6 were obtained by this formulation. Each point in the energy surface plots of Figure 6 is the result of one computation as outlined above. Typical computations involved ~300,000 elements and ran for ~30 hours on a node with 16 cores, 4GB RAM and a clock speed of 3GHz.

**RESULTS**

**Experimental tumor growth model**

Colon cancer cells (LS174T cell line) were incorporated, either as a single-cell suspension or pre-produced tumor spheroids, with 3 cc of liquid agarose before it gelled. The density of tumors and concentration of the agarose gel were controllable. The number of tumors in the gel ranged from one to the number that developed after inoculation of 10,000 cells/cc (an estimated 50% to 70% of embedded "single" cells do not form tumors). Agarose concentrations from 0.3% to 2.0% were used to study tumor growth in gels that were in the range of being just more compliant to just stiffer (0.3% agarose and 0.5% agarose: 0.3 ± 0.2 kPa and 0.7 ± 0.1 kPa) or significantly stiffer (1.0% agarose and 2.0% agarose: 4.0 ± 0.5 kPa and 24 ± 3 kPa) than pre-produced tumor spheroids (0.45 ± 0.03 kPa) [28]. All growth experiments were conducted in 6-well porous-bottomed cell culture inserts and only tumors that were well away from the gel boundaries were measured.

The key parameters measured during growth were tumor size and shape. Tumor size, which increases with time after embedment [8,9], was affected by both tumor density and agarose concentration [29]. Tumor shape, however, was consistently observed to develop to oblate ellipsoidal (semi-axes $a_1 >\approx a_2 > a_3$, Fig. 1) regardless of tumor density. This shape formed within one week after embedment (Fig. S4; Movies S1, S2 in the Supporting Material) and we confirmed that it did not form due to collapse of a necrosed core (Fig. S5). The only parameter affecting this shape development was the agarose elastic modulus: when the agarose gel was stiffer than the pre-produced tumors, the latter grew as oblate ellipsoids.

When the gel's elastic modulus was below that of the pre-produced tumors, approximately spherical tumors grew, with diffuse boundaries (Fig. S3).

**Tumor shape and orientation characterization**

The three-dimensional shape of the tumors was investigated using light sheet fluorescence microscopy (Figs. 1 and 2) and physical sectioning of gels to obtain at least two perpendicular views of individual tumors with a confocal microscope (Fig. S6). Full 3D measurements were made of 11 tumors in 0.5% or 1.0% agarose (Table S1). Eight different null hypotheses were devised to test whether these shapes were oblate ellipsoidal and the Bonferroni adjustment was used to correct for the increased risk of type I errors (false positives, further details in *SI Text*). The sum of the p values for all 8 tests was 0.0064 (Table S2), well under the set statistical significance level for each test (here, p/8 = 0.00625, obtained for p = 0.05). Furthermore, the average 3D aspect ratio ($a_1/a_3$) of tumors in the 1% agarose gel was 2.7 ± 0.3, confirming uniformity in oblate tumor shape.

Three-dimensional reconstructions (Fig. 2b; Movies S3, S4) of light sheet-imaged tumors revealed that the oblate ellipsoidal tumor shape was correlated with a wide range of projected elliptical shapes, from a circle ($a_1 \approx a_2$) to an ellipse of the maximum achievable aspect ratio ($a_1/a_3$), with a range of smaller aspect ratios in between (Fig. 2a). The length of the major axis of the ellipse that is projected by an oblate ellipsoid will always be equal to the length of the largest axis of the oblate ellipsoid itself (the outline of a proof is presented in *SI Text*). Therefore, all possible projected aspect ratios of such an oblate ellipsoid may be calculated by rotating it around one of its major axes (Fig. 3).

Since the oblate tumor shape is robust and its maximum 3D aspect ratio is relatively uniform for tumors grown in 1% agarose gel, we have used the projected aspect ratio as a measure of tumor orientation—defined by rotation around one of its major axes (Fig. 3)—for many tumors. In several experiments with a cell inoculation density of 2500 cells/cc in 1% agarose gel, a maximum aspect ratio of 3 was observed after one month of growth, which is consistent with the average 3D aspect ratio measurement of 2.7 ± 0.3 under the same conditions. The distribution of projected aspect ratios, and therefore tumor rotations, is shown in Fig. 4 for one experiment. After removing the gel and sectioning it to image the tumors from the side, a similar distribution of projected aspect ratios resulted (Fig. 4). This lack of preferential tumor orientation indicates there is no influence of an externally applied field, such as a mechanical stress field, which could bias all the tumor ellipsoidal axes to the same direction.

**Nonlinear elasticity finite element computations**

If an elastic field is associated with the ellipsoidal shapes, such a field must bear a random relation to position since the major axes of the ellipsoidal tumors are randomly oriented with respect to the tumor's position in the gel (Fig. 4). One such example is a tumor growth-induced stress field, which arises in the tumor and gel. If it were unconstrained by the gel, uniform growth would cause a stress-free expansion of the tumor, with possible shape changes, such as of a sphere to an ellipsoid. It is represented exactly by the finite growth tensor $F^g = diag(\alpha_1, \alpha_2, \alpha_3)$. The determinant gives the growth volume ratio: $det(F^g) = \alpha_1\alpha_2\alpha_3$. When constrained by the gel, however, the actual expansion of the tumor differs from $F^g$, and is given by the deformation gradient tensor, $F$. It satisfies the previously introduced elasto-growth decomposition $F=F^e F^g$ [21] where $det(F) < det(F^g)$, and $F^e$ causes the growth stress. The random position, orientation, and growth tensors of the tumors make the growth stress tensor due to each tumor also random in magnitude and orientation.

The stresses in the tumor and gel are reported in Fig. 5 for a spherical tumor and an oblate ellipsoidal tumor ($a_1 = a_2$, $a_1/a_3 = 3$), both growing isotropically ($\alpha_1 = \alpha_2 = \alpha_3$) in gels that are ten times stiffer ($\mu^{tum}/\mu^{gel} = 1/10$), matching our experiments for tumors in 1% agarose gel. Growth leads to a compressive stress inside the tumor when it is constrained by the gel. For a spherical tumor growing isotropically, the compressive (negative) stress is uniform in every direction ($\sigma_{11} = \sigma_{22} = \sigma_{33} = \sigma_{sph}$). However, for an ellipsoidal tumor growing isotropically, the stress components within the tumor are no longer equal: For an oblate ellipsoidal tumor ($a_1 = a_2$, $a_1/a_3 > 1$), the compressive stresses along the $a_1$ and $a_2$ axes satisfy $\sigma_{11} = \sigma_{22}$, and are greater than that along the short axis, $|\sigma_{11}| > |\sigma_{33}|$. For a growth tensor $\boldsymbol{F}^g$, these compressive stress magnitudes can be computed to be continuous at the tumor-gel boundary, and to decay outside the tumor. Tensile normal stress components are induced in the hydrogel tangential to the tumor-gel boundary, but no cracking of the hydrogel was detected as a result (Fig. S2, for further details see Supporting Material).

We computed the elastic free energy, $\psi$, of the tumor-gel system for large isotropic growth of the form $\boldsymbol{F}^g = diag(5^{1/3}, 5^{1/3}, 5^{1/3})$, $det\boldsymbol{F}^g = 5$. A total of 576 boundary value problems were solved, each with different ratios $a_1/a_3$ and $a_2/a_3$. The initial ellipsoidal volume was fixed by imposing $a_1 a_2 a_3 = 1.25 \times 10^5$ $\mu m^3$. Results have been plotted in Fig. 6 for the elastic free energy $\psi$, versus $a_1/a_3$ and $a_2/a_3$ in the interval $1 \leq a_1/a_3, a_1/a_3 \leq 3$. Matching our experiments with tumors grown in 1% agarose hydrogel, Fig. 6a summarizes the elastic free energy results for $\mu^{tum}/\mu^{gel} = 1/10$. The free energy is highest in the neighborhood of approximately spherical tumors, $a_1/a_3 \sim 1$—$1.5$ and $a_2/a_3 \sim 1$—$1.5$, and lowest in the limit of oblate ellipsoidal tumors: $a_1/a_3, a_2/a_3 \sim 3$—the shape most often observed in our experiments. The energy surface in Fig. 6a also shows that for a fixed value of $a_1/a_3$ ($a_2/a_3$) the energy is minimized for the largest $a_2/a_3$ ($a_1/a_3$), although this variation is more gradual. This result suggests that, if $\mu^{tum}/\mu^{gel} < 1$, the reason for the predominance of the oblate ellipsoidal tumor shape could be associated with the fact that this shape minimizes the elastic free energy.

I*n vivo* soft tissue tumors, however, are observed to be stiffer than the surrounding tissue [2,30,31]. While we have not been able to replicate this regime in our experiments, we approach it in the case of 0.3% agarose gels, which corresponds to $\mu^{tum}/\mu^{gel} <\approx 1$. As shown in Fig. 6b, the elastic free energy landscape is much flatter than in Fig. 6a, because the energy differences are much lower. This suggests that configurations closer to spherical are subject to a less stringent free energy penalty. (Note: These results—both for the stress field and the elastic free energy—remain valid even if linearized, infinitesimal strain elasticity is used following Mura's [32] treatment of Eshelby's inclusion problem [33,34]. While the formulas are more tractable and the results may be directly plotted without the need for finite element computations, the validity of the calculations is doubtful because of the finite strains implied by large tumor growth ratios.) This finding corroborates the observation in our experiments in this case for which the tumors remain closer to spherical (Fig. S3).

**DISCUSSION**

Monitoring displacements in the hydrogel surrounding ellipsoidal tumors using co-embedded fluorescent micro-beads, Cheng and co-workers [7] sought to explain the observed symmetry-breaking that leads to ellipsoidal tumors. The strain fields computed from the micro-bead displacements were interpreted as showing the compressive stress to be greater along the minor axis of the ellipsoid. By correlating these fields with tumor shape and Caspase-3 activity, the authors concluded that mechanical stress was causing a higher fraction of cell death along the minor axis and driving the tumor to grow in the corresponding ellipsoidal shape. The authors did not, however, identify the origin of the stress in their system.

We were unable to determine any correlation between tumor orientation and position in the gel (Fig. 4). As we have observed above, if the ellipsoidal shape were associated with an elastic field, that field could not be correlated with position in the gel because of the observed random distribution of ellipsoidal tumor axis orientations. We also have noted that the stress field caused by the growth (expansion) of the tumor in the gel is an example of such an elastic field. Nonlinear elasticity, however, runs contrary to the conclusions of Cheng and co-workers [7]. To demonstrate this, we have carried out nonlinear elasticity computations to show that the oblate ellipsoidal growth of a tumor induces compressive stress components along its major axes that are greater in magnitude than the component along the minor axis when measured just outside the tumor (Fig 5). This result is also obtained with linearized, infinitesimal elasticity. If compressive stress suppresses growth by signaling different cell growth and/or death rates, the major axes ought to suffer this suppression more sharply than the minor axis. Such a sequence of events would favor the spherical shape, which clearly does not happen in our experimental studies. The experimental evidence suggests that the growth-induced compressive stresses, which are higher along the major axes, do not suppress the growth of ellipsoidal tumors. Therefore, such shapes must be controlled by another quantity.

Over the course of our studies, we computationally tested several mechanisms that could drive tumor growth into an oblate ellipsoidal shape. These included (a) growth along a compliant layer, (b) the rise of a highly proliferative subpopulation of cells, (c) stress-driven migration of cells, and (d) elastic free energy minimization. Through extensive tumor growth modeling based on our past work [21-23] and that of Casciari et al. [35], we concluded that the mechanisms in (a)—(c) could account for small perturbations of the tumor shape toward an ellipsoid, but not for the strongly ellipsoidal shapes that we have observed (see reference [23] in this regard). The study of the elastic free energy of the tumor-gel system, however, has proved more promising. Indeed, for the case of a more compliant tumor in a stiff matrix ($\mu^{tum}/\mu^{gel} = 1/10$) a steep energy surface results in which the elastic free energy is significantly minimized for oblate ellipsoidal tumor shapes (Fig. 6a). Although the oblate ellipsoidal shape is still the energy-minimizing shape if the tumor and gel have equal shear moduli, the free energy penalty for remaining closer to a sphere is just about a fifth of the penalty for the case $\mu^{tum}/\mu^{gel} = 1/10$.

We confirmed that the orientation of the semi-axes of the ellipsoidally growing tumor had no influence on the total strain energy of the tumor-gel system, as long as the gel boundaries were sufficiently distant. We also confirmed that the location of the tumor relative to the gel boundaries had negligible effect on the elastic free energy. For a tumor of a given shape the elastic free energy varied by less than 1% with as its position in the gel was varied. This included positions that placed the tumor edge within one radius, or one semi-major axis of the gel boundary. This result is expected also from linearized elasticity. Given these computational results, we suggest that any tendency for tumors to grow parallel to gel boundaries is due to mechanical contact resulting in a force that deforms the growing tumor into such a shape. We did not report contact mechanics computations of the elastic free energy for tumors growing in contact with the gel boundary. The elastic free energy in such contact mechanics computations is very sensitive to the gap function chosen to impose contact. As noted above, our experimental set up was not conducive to precise measurement of tumor proximity to gel boundaries, because of optical distortion at the boundaries. We therefore are not in a position to make a rigorous report on the boundary effect on the basis of our experiments. However, as also stated above, all tumors reported in our experiments were far from the gel boundaries.

We note that the qualitative trends observed with the formulation of tumor growth as a problem of finite strain nonlinear elasticity are reproduced with infinitesimal, linearized elasticity following Refs [32-34]. However such calculations using the linearized theory lack the physical consistency and quantitative accuracy of the full nonlinear theory.

Extrapolating the above results to *in vivo* observations, we propose that the observed tendency toward ellipsoidal solid tumors [10-13,36] results from the lower elastic free energy of the ellipsoidal shape in the early prevascular stage of growth when they are compliant relative to the surrounding tissue. (Note: While we are not aware of any observations of pre-vascular tumors in vivo, we note that the pre-vascular tumors in our study have moduli in the range of 0.3-0.7 kPa, which is at the lower extreme of soft tissues as reported by Yu et al. [2].) Subsequently, as these tumors grow larger and vascularize, and their microenvironment becomes more fibrotic, the elastic modulus of the tumor tissue may exceed that of the surrounding tissue [2]. In this regime of tumor *versus* surrounding tissue stiffness, spherical tumors are not subject to as stringent an energy penalty, and therefore are more likely to be found. However, with early prevascular tumors being more compliant than their microenvironment, the penalization of the spherical shape at this stage ensures that it is the ellipsoidal tumors, especially the oblate ellipsoidal ones, that are favored. Our nonlinear elasticity computations have further shown that there is a stringent energy penalty against shape changes that could be effected by non-diagonal growth tensors: $\bm{F}^g = diag(\alpha_1, \alpha_2, \alpha_3)$, where at least one of $\alpha_1$, $\alpha_2$ and $\alpha_3$ is different from the others. Because of these energy-dependent mechanisms, ellipsoidal tumors are favored over other configurations, and are observed more often, across all tumor-gel stiffness ratios.

Our findings do not suggest what mechanisms of growth could actually suppress the high-energy shapes. Indeed, our methods, experimental and theoretical, are not designed to answer this question. One possibility is that, for tumor-gel systems with a greater elastic free energy, the mechanical work performed by a growing tumor to store this energy drains the biochemical free energy of the cells in the early stages of tumor development. This hypothesis would have to be tested against the complexities of cancer metabolism [37] before reaching a firmer conclusion.  Because agarose gels are biochemically inert, mechanical interactions play a more dominant role in determining the shapes of tumors in our study, by design. *In vivo*, of course, there are other biochemical drivers of tumor growth. However, the results of this study would suggest that minimization of elastic free energy is an important driver of tumor shape and, due to significant reporting of ellipsoidal tumors, possibly plays a significant role *in vivo* also.

The importance of this work may prove to be that tumor shape plays a role in determining the potential for malignancy. Once it has a well-developed necrotic core, a tumor can be modeled as an internally pressurized thin shell carrying meridional and azimuthal stress resultants in its wall, which is composed of viable cells. These stress resultants are known from growth models to control shape instabilities of the tumor wall [38-40], leading ultimately to its breakdown. Thin shell theory [41] shows that, while spherical shells have equal azimuthal and meridional stresses, both being uniform over the shell, ellipsoidal shells have non-uniform fields with extrema at the poles of the ellipsoid. These points on an ellipsoidal tumor could therefore be critical sites for the surface instability, breakdown of the wall, and potential cell escape leading to malignancy.

**ACKNOWLEDGEMENTS**

Financial support was provided by the Alexander von Humboldt and Max Planck Societies. Assistance acquiring light sheet fluorescence microscope images was provided in the lab of Prof. H. Schneckenburger by S. Schickinger.

# REFERENCES

1. Kass L, Erler JT, Dembo M, Weaver VM (2007) Mammary epithelial cell: Influence of extracellular matrix composition and organization during development and tumorigenesis. Int J Biochem Cell B 39: 1987–1994.

2. Yu H, Mouw JK, Weaver VM (2011) Forcing form and function: biomechanical regulation of tumor evolution. Trends Cell Biol 21: 47–56.

3. Shieh AC (2011) Biomechanical forces shape the tumor microenvironment. Ann Biomed Eng 39: 1379–1389.

4. Paszek MJ, Zahir N, Johnson KR, Lakins JN, Rozenberg GI, et al. (2005) Tensional homeostasis and the malignant phenotype. Cancer Cell 8: 241–254.

5. Tse JM, Cheng G, Tyrrell JA, Wilcox-Adelman SA, Boucher Y, et al. (2012) Mechanical compression drives cancer cells toward invasive phenotype. Proc Natl Acad Sci USA 109: 911–916.

6. Demou ZN (2010) Gene expression profiles in 3D tumor analogs indicate compressive strain differentially enhances metastatic potential. Ann Biomed Eng 38: 3509–3520.

7. Cheng G, Tse J, Jain RK, Munn LL (2009) Micro-environmental mechanical stress controls tumor spheroid size and morphology by suppressing proliferation and inducing apoptosis in cancer cells. PLoS ONE 4: e4632.

8. Helmlinger G, Netti PA, Lichtenbeld HC, Melder RJ, Jain RK (1997) Solid stress inhibits the growth of multicellular tumor spheroids. Nat Biotechnol 15: 778–783.

9. Koike C, McKee TD, Pluen A, Ramanujan S, Burton K, et al. (2002) Solid stress facilitates spheroid formation: potential involvement of hyaluronan. Brit J Cancer 86:

947–953.

10. Dassios G, Kariotou F, Tsampas MN, Sleeman BD (2012) Mathematical modelling of avascular ellipsoidal tumour growth. Q Appl Math 70: 1–24.

11. Güth U, Brenckle D, Huang DJ, Schötzau A, Viehl CT, et al. (2009) Three-dimensional pathological size assessment in primary breast carcinoma. Breast Cancer Res Tr 116: 257–262.

12. Wapnir I, Wartenberg D, Greco R (1996) Three dimensional staging of breast cancer. Breast Cancer Res Tr 41: 15–19.

13. Costantini M, Belli P, Lombardi R, Franceschini G, Mulè A, et al. (2006) Characterization of solid breast masses: use of the sonographic breast imaging reporting and data system lexicon. J Ultrasound Med 25: 649–659.

14. Carlsson G, Gullberg B, Hafström L (1983) Estimation of liver tumor volume using different formulas - an experimental study in rats. J Cancer Res Clin Oncol 105: 20–23.

15. Dachman AH, MacEneaney PM, Adedipe A, Carlin M, Schumm LP (2001) Tumor size on computed tomography scans: is one measurement enough? Cancer 91: 555–560.

16. Brenner MW, Holsti LR, Perttala Y (1967) The study by graphical analysis of the growth of human tumours and metastases of the lung. Brit J Cancer 21: 1–13.

17. Folkman J, Hochberg M (1973) Self-regulation of growth in 3 dimensions. J Exp Med 138: 745–753.

18. Mueller-Klieser W (1987) Multicellular Spheroids - A review on cellular aggregates in cancer-research. J Cancer Res Clin Oncol 113: 101–122.

19. Kelm JM, Timmins NE, Brown CJ, Fussenegger M, Nielsen LK (2003) Method for generation of homogeneous multicellular tumor spheroids applicable to a wide variety of cell types. Biotechnol Bioeng 83: 173–180.

20. Bruns T (2012) Preparation strategy and illumination of three-dimensional cell cultures in light sheet–based fluorescence microscopy. J Biomed Opt 17: 101518.

21. Garikipati K, Arruda EM, Grosh K, Narayanan H, Calve S (2004) A continuum treatment of growth in biological tissue: the coupling of mass transport and mechanics. J Mech Phys Solids 52: 1595–1625.

22. Narayanan H, Verner SN, Mills KL, Kemkemer R, Garikipati K (2010) In silico estimates of the free energy rates in growing tumor spheroids. J Phys:-Condens Mat 22: 194122.

23. Rudraraju S, Mills K, Kemkemer R, Garikipati K (2013) Multiphysics Modeling of Reactions, Mass Transport and Mechanics of Tumor Growth. In: Holzapfel GA, Kuhl E, editors. Computer Models in Biomechanics. Dordrecht: Springer Netherlands. pp. 293–303.

24. Ambrosi D, Ateshian GA, Arruda EM, Cowin SC, Dumais J, et al. (2011) Perspectives on biological growth and remodeling. J Mech Phys Solids 59: 863–883.

25. Garikipati K, Rao VS (2001) Recent Advances in Models for Thermal Oxidation of Silicon. J Comput Phys 174: 138–170.

26. Flory PJ (1961) Thermodynamic relations for high elastic materials. T Faraday Soc 57: 829–838.

27. Bangerth W, Hartmann R, Kanschat G (2007) deal.II---A general-purpose object-

oriented finite element library. ACM Trans Math Softw 33: 24.

28. Mills KL, Garikipati K, Kemkemer R (2011) Experimental characterization of tumor spheroids for studies of the energetics of tumor growth. Int J Mater Res 102: 889–895.

29. Mills KL, Kemkemer R, Rudraraju S, Garikipati K. Growth of prevascular tumors in soft hydrogels is mechanically coupled. Manuscript in progress.

30. Dewall RJ, Bharat S, Varghese T, Hanson ME, Agni RM, et al. (2012) Characterizing the compression-dependent viscoelastic properties of human hepatic pathologies using dynamic compression testing. Phys Med Biol 57: 2273–2286.

31. Samani A, Bishop J, Luginbuhl C, Plewes D (2003) Measuring the elastic modulus of ex vivo small tissue samples. Phys Med Biol 48: 2183–2198.

32. Mura T (1987) Micromechanics of defects in solids. 2nd ed. Dordrecht: Kluwer Academic Publishers.

33. Eshelby JD (1957) The determination of the elastic field of an ellipsoidal inclusion, and related problems. Proc R Soc Lon Ser-A 241: 376–396.

34. Eshelby JD (1959) The elastic field outside an ellipsoidal inclusion. Proc R Soc Lon Ser-A 252: 561–569.

35. Casciari JJ, Sotirchos SV, Sutherland RM (1992) Variations in tumor-cell growth-rates and metabolism with oxygen concentration, glucose-concentration, and extracellular pH. J Cell Physiol 151: 386–394.

36. Fornage BD, Lorigan JG, Andry E (1989) Fibroadenoma of the breast: sonographic appearance. Radiology 172: 671–675.

37. Weinberg RA (2007) The Biology of Cancer. 1st ed. Garland Science.

38. Chaplain MA, Sleeman BD (1993) Modelling the growth of solid tumours and incorporating a method for their classification using nonlinear elasticity theory. J Math Biol 31: 431–473.

39. Greenspan HP (1976) On the growth and stability of cell cultures and solid tumors. J Theor Biol 56: 229–242.

40. Landman KA, Please CP (2001) Tumour dynamics and necrosis: surface tension and stability. IMA J Math Appl Med 18: 131–158.

41. Flügge W (1960) Stresses in Shells. Berlin: Springer-Verlag.

**FIGURE LEGENDS**

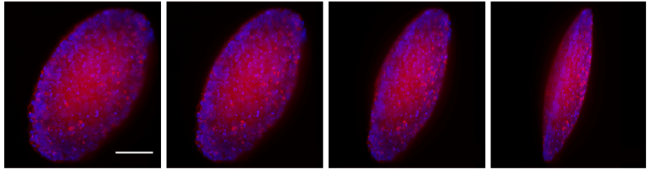

**Figure 1. 3D rendering of an oblate ellipsoidal tumor.** The 3D rendering, created using images taken with a light sheet fluorescence microscope, is rotated about the vertical axis in this sequence of images. Blue: Hoechst-stained nuclei; Red: E-cadherin. Scale bar is 250 μm.

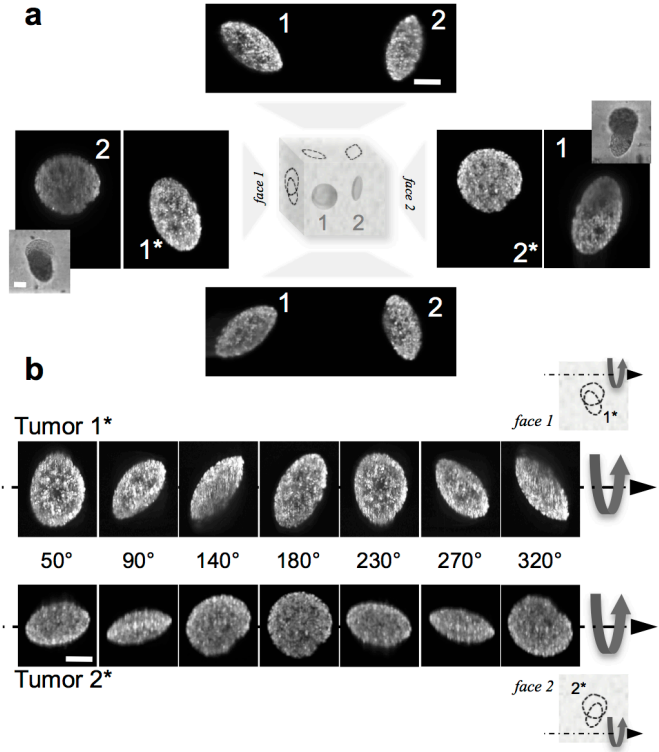

**Figure 2. The various projections of oblate ellipsoidal tumors. a.** Projections (maximum intensity) of two oblate ellipsoidal grown in 0.5% agarose gel and imaged with a light sheet fluorescence microscope from four different directions. **b.** Rotational sequences of the 3D renderings. The 0° orientation is marked with an asterisk in **a**. Complete sequences are available as Movies S3 & S4. Fluorescence signal is from Hoechst-stained nuclei. Scale bars are all 50 μm.

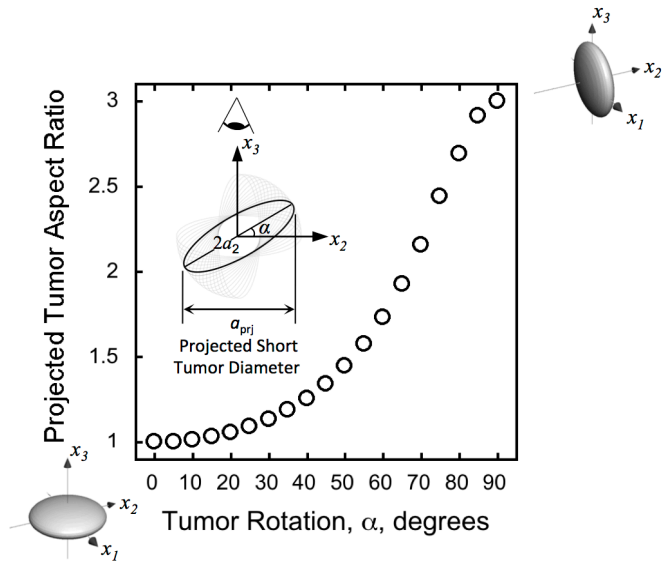

**Figures 3. Projected aspect ratio and oblate ellipsoidal rotation.** The relationship between tumor rotation angle, $\alpha$, and projected aspect ratio, $a_1/a_{prj}$, as an oblate ellipsoid with maximum aspect ratio of 3 ($a_1/a_3 = 3$) is rotated about its $a_1$ ($x_1$) axis.

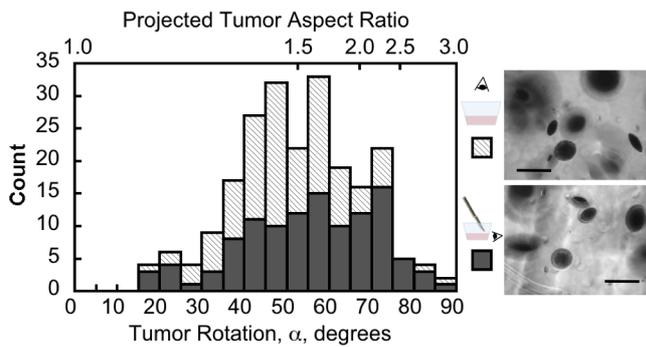

**Figure 4. Oblate ellipsoidal tumors have a similar distribution of orientations regardless of viewing direction.** The tumor rotation angle was calculated from the projected aspect ratio (Fig. 3) for tumors grown in 1% agarose for 30 days. Top image: observation perpendicular to the plane of the cell-culture well, Bottom image: side view perpendicular to a physical cross-section of the gel made with a scalpel blade (out-of-focus cut marks are visible in the gel) Scale bars are 200 μm.

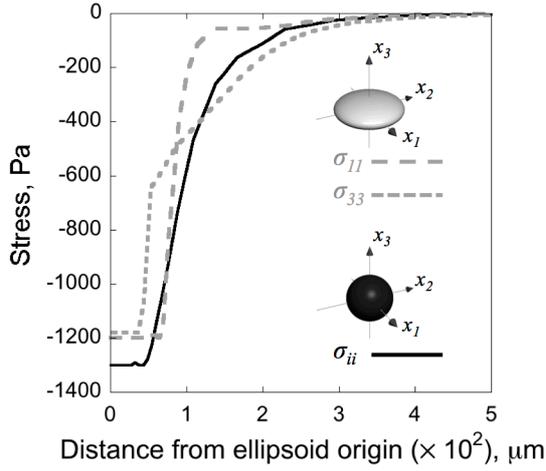

**Figure 5. The stress field created by tumor growth when $\mu^{tum}/\mu^{gel} = 1/10$.** The growth volume ratio det$F^g = 10$. All normal stress components, $\sigma_{ii}$ (———), are equal in a spherical tumor of initial radius 50 μm. The maximum compressive stress is -1300 Pa. However, in an ellipsoidal tumor with axes $a_1 = a_2 = 3a_3 = 72$ μm, the $\sigma_{11}$ (– – –) component, has a maximum compressive stress of -1200 Pa, and the $\sigma_{33}$ (········) component has a maximum compressive stress of -1180 Pa.

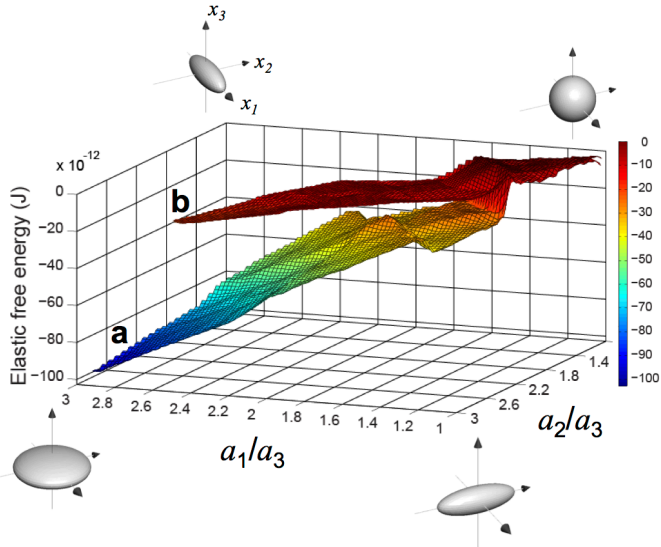

**Figure 6. Elastic free energy landscapes of the tumor-gel system.** The landscapes are plotted *versus* the tumor's ellipsoidal axes ratios $a_2/a_3$ and $a_1/a_3$. The origin has been shifted, so that the maximum energy is zero in each case: **(a)** For $\mu^{tum}/\mu^{gel} = 1/10$. The oblate shapes ($a_2/a_3 = a_1/a_3 = 3$) are the low energy states. This surface has been generated from 576 separate nonlinear elasticity computations of tumors of aspect ratios $a_1/a_3$ and $a_2/a_3$ varying between 1 and 3, growing in a gel. **(b)** The same as (a), but for $\mu^{tum}/\mu^{gel} = 1$. Note the flatness of the landscape relative to **(a)**. Spherical shapes are not penalized strongly for $\mu^{tum}/\mu^{gel} = 1$, but are strongly penalized for $\mu^{tum}/\mu^{gel} = 1/10$.

## SUPPORTING INFORMATION

**Figure S1. Three examples of a plane intersecting an oblate ellipsoid.** The direction of projection, with unit vector, *e*, is drawn from an arbitrary point of observation through the centroid of the oblate ellipsoid. The plane, *P*, is normal to *e* and contains the centroid of the oblate ellipsoid. The ellipse that is created by the intersection of any such plane with the oblate ellipsoid has as its larger axis the major axes of the oblate ellipsoid: $2a_1 = 2a_2$.

**Figure S2. Cracks in the gel are not associated with tumor growth. (a)** An oblate ellipsoidal tumor with its minor axis ($a_3$) oriented in the projection view grows from a major diameter of 160 mm and aspect ratio of 1.2 to a major diameter of 920 μm and aspect ratio of 2.8 over the course of 10 days. The scale bar is 200 μm. **(b)** Above and to the left of the tumor in this image, the bottom half of a crack in the hydrogel is visible that is not associated with a tumor growing inside of it. Note that the hydrogel is cleaved to the upper left of the crack boundary in this image. The scale bar is 500 μm.

**Figure S3. Tumors grown in 0.3% agarose. Left Column:** Phase contrast images of 18-day-old tumors taken with a 5x objective. Scale bar is 500 μm. **Right Column:** Projections at two perpendicular orientations of a 15-day-old tumor. The fluorescent signal is from the Hoechst-stained nuclei. The scale bar is 50 μm. The tumor shape in the softest gels is often more diffuse and approximately spherical.

**Figure S4. Individual tumor development.** To determine the time frame over which the oblate ellipsoidal shape developed, individually embedded, pre-produced tumor spheroids were followed as they grew in the agarose gel. In such an experiment, it was determined that within one week after embedment the ellipsoidal shape was detectable and the orientation of the tumor in the gel was fixed. Shown here is the time course development of the projected tumor aspect ratio of 11 different individual tumors (sparsely embedded). Diamond symbols mark the aspect-ratio measurements for tumors whose final orientation projected narrow ellipses, solid circles for tumors whose final orientation projected near-circles, and open circles for tumors whose final orientation projected wider ellipses. Phase-contrast images show the projections from which the aspect ratios were measured for the specific tumors and time points circled in grey on the plot. Scale bars are 500 μm.

**Figure S5. In the ellipsoidal cross-section of a three-day-old tumor, a necrotic core is not detectable.** It was confirmed that the tumors took on the oblate ellipsoidal shape before the observation of a central necrotic core, ruling out the possibility that this shape formed due to mechanical collapse of a necrosed core. Three-dimensional projection of 13 image slices, taken 2 μm apart in the mid-section of an oblate ellipsoidal tumor. The section is shown here tilted about the vertical axis of the image by 10 degrees. Nuclei (blue) and actin network (red) of LS174T cells embedded in 1% agarose hydrogel. Scale bar is 30 μm.

**Figure S6. Perpendicular views of 5 tumors grown in 1% agarose.** Due to scattering regular confocal microscopy is not effective at capturing entire 3D tumor shapes. However, by carefully sectioning the gel in the region of a tumor, two approximately perpendicular views may be imaged of the tumors. Here are some examples. The fluorescent signal is provided by Hoechst-stained nuclei. The scale bars are 100 μm long.

**Movies S1 & S2.** Time lapse progression of individually embedded tumors that developed elliptical and approximately circular cross-sections, respectively. Scale bars are 500 μm.

**Movies S3 & S4.** 3D renderings of Tumor 1 and Tumor 2 (Figure 2 in the main paper), respectively.

# Elastic free energy drives the shape of prevascular solid tumors

## Supporting Information

K. L. Mills, Ralf Kemkemer, Shiva Rudraraju, and Krishna Garikipati

correspondence to: mills@is.mpg.de or krishna@umich.edu

**Table of Contents**

Statistical testing on measured 3D tumor dimensions

| Tumor designation | | | Tumor diameter, μm | | |
|---|---|---|---|---|---|
| | | | $2a_1$ | $2a_2$ | $2a_3$ |
| | 0.5% | 1) Fig. 2 Tumor 1 | 289 | 248 | 117 |
| | | 2) Fig. 2 Tumor 2 | 247 | 227 | 114 |
| | | 3) R005 | 91 | 74 | 39 |
| | 1.0% | 4) Fig. 1 | 371 | 358 | 138 |
| | | 5) R002 | 207 | 183 | 80 |
| | | 6) R004 | 514 | 507 | 157 |
| | | 7) Fig. S6 Tumor A | 445 | 442 | 140 |
| | | 8) Fig. S6 Tumor B | 315 | 308 | 123 |
| | | 9) Fig. S6 Tumor C | 179 | 176 | 74 |
| | | 10) Fig. S6 Tumor D | 283 | 264 | 120 |
| | | 11) Fig. S6 Tumor E | 517 | 427 | 181 |

**Table S1.** 3D measurements of the tumor axes $2a_1$, $2a_2$ and $2a_3$ from 11 different tumors in 0.5% and 1.0% agarose hydrogels. The second column is the tumor designation, which, when included in the paper or this Supplementary Materials document, is referred to by its caption number.

Definitions of Null Hypotheses:
*The tumor shape is not oblate ellipsoidal*
(a)   1) difference between $a_1$ and $a_2$, $(a_1-a_2)/a_1$, is greater than 0.15 ($\mu_{0,1} > 0.15$) AND
      2) oblateness, $f = (a_{ave.1,2} - a_3)/a_{ave.1,2}$, is less than 0.5 ($\mu_{0,2} < 0.5$) (i.e., the long axes are less than 2 times as long as the short axis)

(b)   3) $a_1/a_3 < 2$ ($\mu_{0,3} < 2$) AND
      4) $a_2$ is closer to $a_3$ than to $a_1$, $(a_1-a_2)/(a_2-a_3) >1$ ($\mu_{0,4} > 1$)

*The tumor shape is a spheroid*
      5) no difference between $a_1$ and $a_2$ ($\mu_{0,5} = 0$) AND
      6) oblateness, $f$, is equal to 0 ($\mu_{0,6} = 0$)

*The tumor shape is a general ellipsoid ($a_2$ is half-way between $a_1$ and $a_3$)*
      7) difference between $a_1$ and $a_2$, $(a_1-a_2)/a_1$ is 0.5 ($\mu_{0,7} = 0.5$) AND
      8) oblateness, $f$, is 0.75 ($\mu_{0,8} = 0.75$)

| Null Hypothesis, $n_i$ | Average, $\bar{x}$ | Standard deviation, $\sigma$ | $t = \dfrac{\bar{x} - \mu_0}{\sigma/\sqrt{n}}$ | p-value (*two-tailed) |
|---|---|---|---|---|
| 1) $\mu_{0,1} > 0.15$ | 0.079 | 0.068 | -3.493 | 0.003 |
| 2) $\mu_{0,2} < 0.5$ | 0.597 | 0.057 | 5.635 | 0.00018 |
| 3) $\mu_{0,3} < 2$ | 2.629 | 0.354 | 5.888 | 0.000077 |
| 4) $\mu_{0,4} > 1$ | 0.171 | 0.163 | -16.898 | $5.5 \times 10^{-9}$ |
| 5) $\mu_{0,5} = 0$ | 0.079 | 0.068 | 3.841 | 0.003* |
| 6) $\mu_{0,6} = 0$ | 0.597 | 0.057 | 34.724 | $9.3 \times 10^{-12}$* |
| 7) $\mu_{0,7} = 0.5$ | 0.079 | 0.068 | -20.605 | $1.6 \times 10^{-9}$* |
| 8) $\mu_{0,8} = 0.75$ | 0.597 | 0.057 | -8.909 | $4.5 \times 10^{-6}$* |

**Table S2.** Statistical test values, by a Student's T-test, for the null hypotheses listed above. The Bonferroni adjustment gives individual significance values for each test of 0.00625 when a total significance value of 0.05 is considered.

The projection of an oblate ellipsoid always contains its major axes

The projection of an oblate ellipsoid is the ellipse that results from the intersection of a plane—oriented normal to the projection direction—as it is passed through the oblate ellipsoid. When this plane shares the centroid with the oblate ellipsoid (Fig. S1), the major axis of the ellipse will be equal in length to the major axes ($2a_1 = 2a_2$) of the oblate ellipsoid, no matter what the projection direction is. Since no other intersection contains a larger axis, this means that the major axis of any projection of an oblate ellipsoid is equal to the major axes of the oblate ellipsoid.

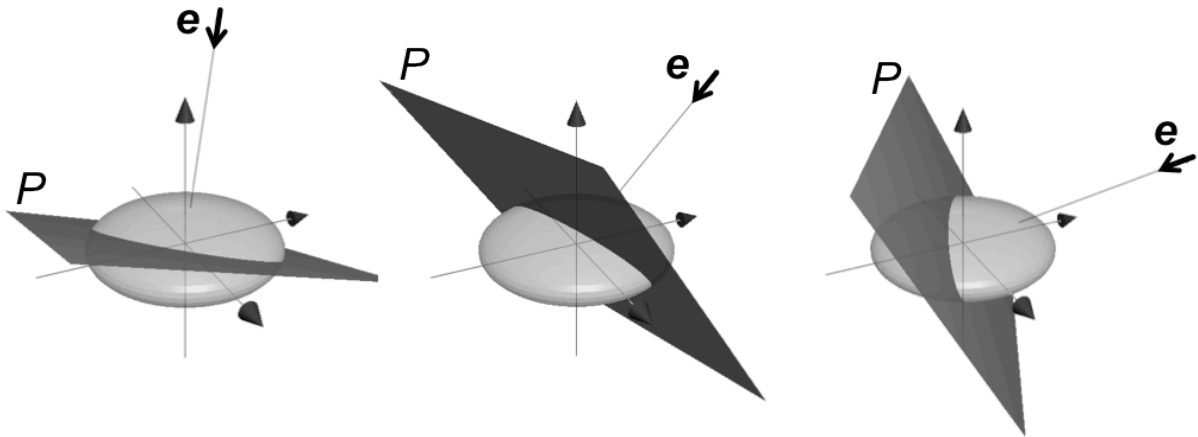

**Figure S1.** Three examples of a plane intersecting an oblate ellipsoid. The direction of projection, with unit vector, $e$, is drawn from an arbitrary point of observation through the centroid of the oblate ellipsoid. The plane, $P$, is normal to $e$ and contains the centroid of the oblate ellipsoid. The ellipse that is created by the intersection of any such plane with the oblate ellipsoid has as its larger axis the major axes of the oblate ellipsoid: $2a_1 = 2a_2$.

Crack initiation in the agarose hydrogels

The growth strain additionally leads to a normal stress in the gel tangential to the tumor-gel boundary: $\sigma_{33}$ along the $x_1$ direction and $\sigma_{11}$ along the $x_3$ direction (26). Notably, these are discontinuous at the tumor-gel boundary, going from compressive within the tumor to tensile in the gel.

Cheng and co-workers (7) suggested that the tensile $\sigma_{33}$ stress at the tip of the major axis of the oblate ellipsoid induces cracking of the agarose hydrogel. We have not found such cracks in this study (Fig. S2a). More often, cracks in the gel, which take on an ellipsoidal cross-section in the plane of the cell culture well, were seen immediately following agarose gelation; and, at later time points in the experiment, they were not associated with a tumor (Fig. S2b).

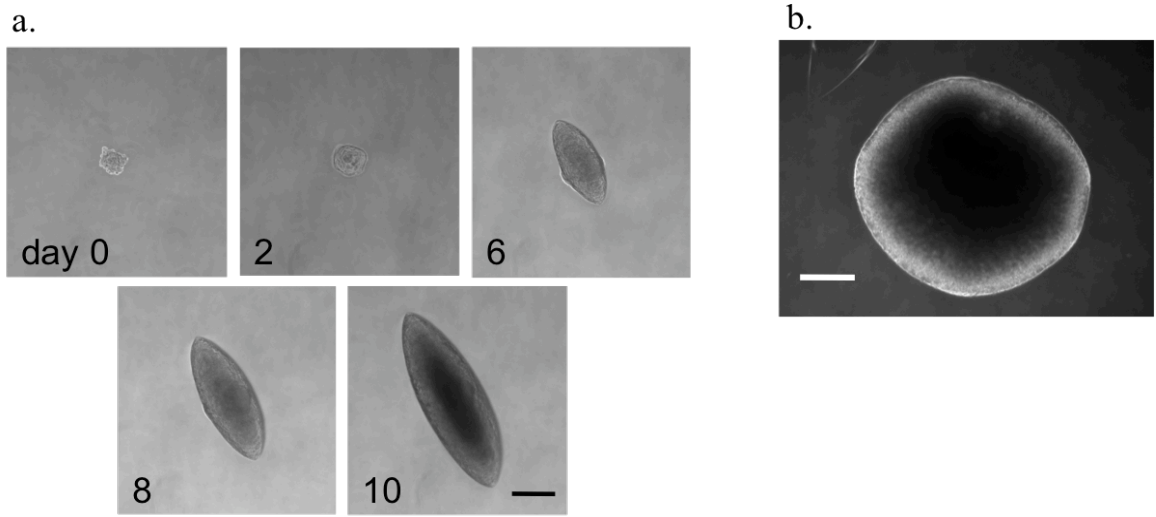

**Figure S2. Cracks in the gel are not associated with tumor growth. (a)** An oblate ellipsoidal tumor with its minor axis ($a_3$) oriented in the projection view grows from a major diameter of 160 mm and aspect ratio of 1.2 to a major diameter of 920 µm and aspect ratio of 2.8 over the course of 10 days. The scale bar is 200 µm. **(b)** Above and to the left of the tumor in this image, the bottom half of a crack in the hydrogel is visible that is not associated with a tumor growing inside of it. Note that the hydrogel is cleaved to the upper left of the crack boundary in this image. The scale bar is 500 µm.

The cracks were sometimes associated with bubbles that were incorporated in the agarose before gelation. When not associated with a bubble, the cracks may have formed during the gelation process itself. If liquid evaporates during this time, the gel will contract. Since the gel is bonded to the walls of the cell-culture well, a tensile stress would develop in the agarose hydrogel in the plane of the cell culture well. Depending on its magnitude, the stress may be large enough to produce a crack in the gel in the orientation that is experimentally observed.

## Author Contributions

KLM, KG, and RK conceived the research and designed the experiments. KLM performed the experiments. KLM analyzed and KLM, KG, and RK interpreted the data. KLM and KG performed the elasticity calculations. KG and SR performed the computational modeling. KLM and KG wrote the paper.

**Supplemental Figures**

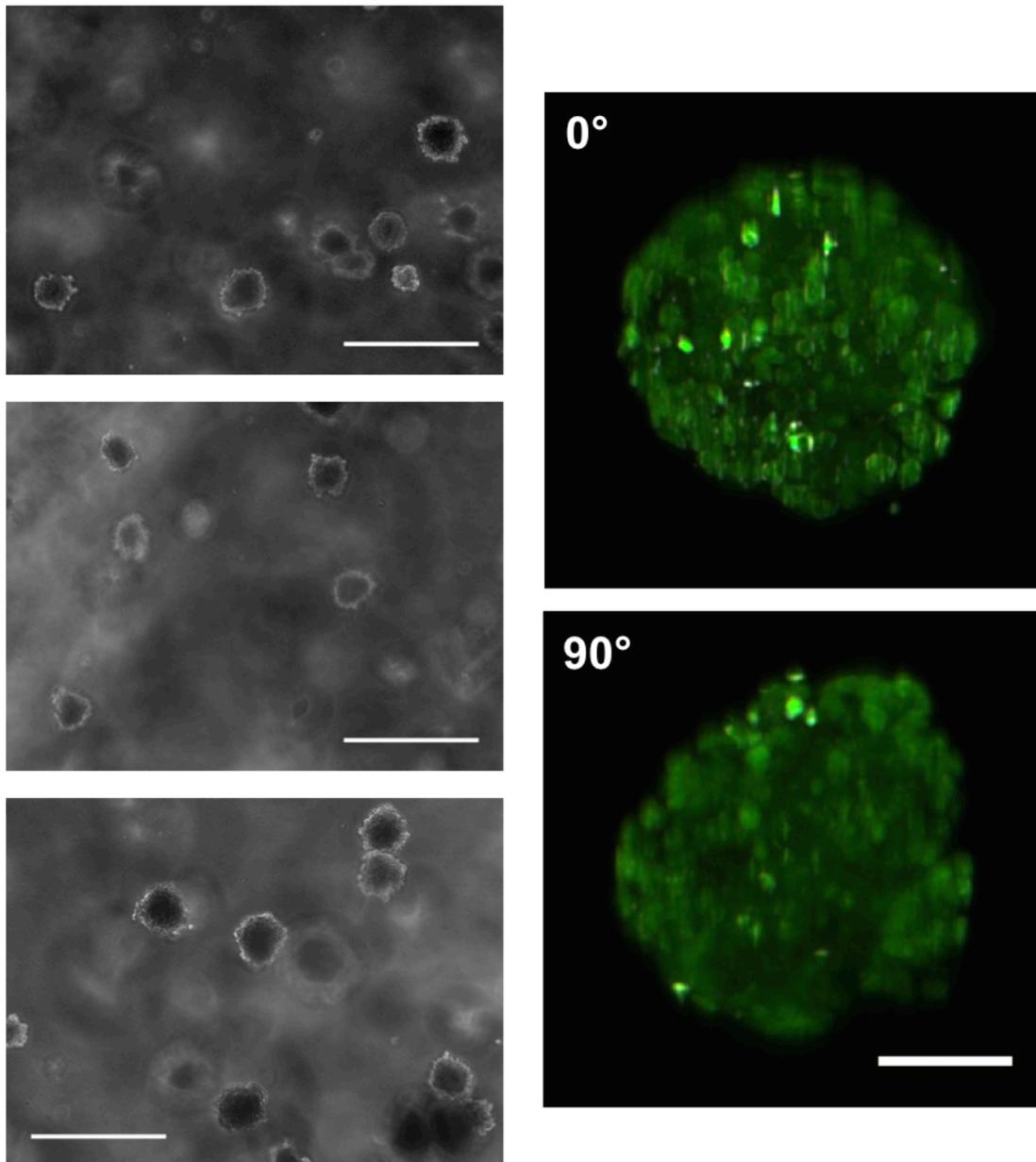

**Figure S3. Tumors grown in 0.3% agarose. Left Column:** Phase contrast images of 18-day-old tumors taken with a 5x objective. Scale bar is 500 μm. **Right Column:** Projections at two perpendicular orientations of a 15-day-old tumor. The fluorescent signal is from the Hoechst-stained nuclei. The scale bar is 50 μm. The tumor shape in the softest gels is often more diffuse and approximately spherical.

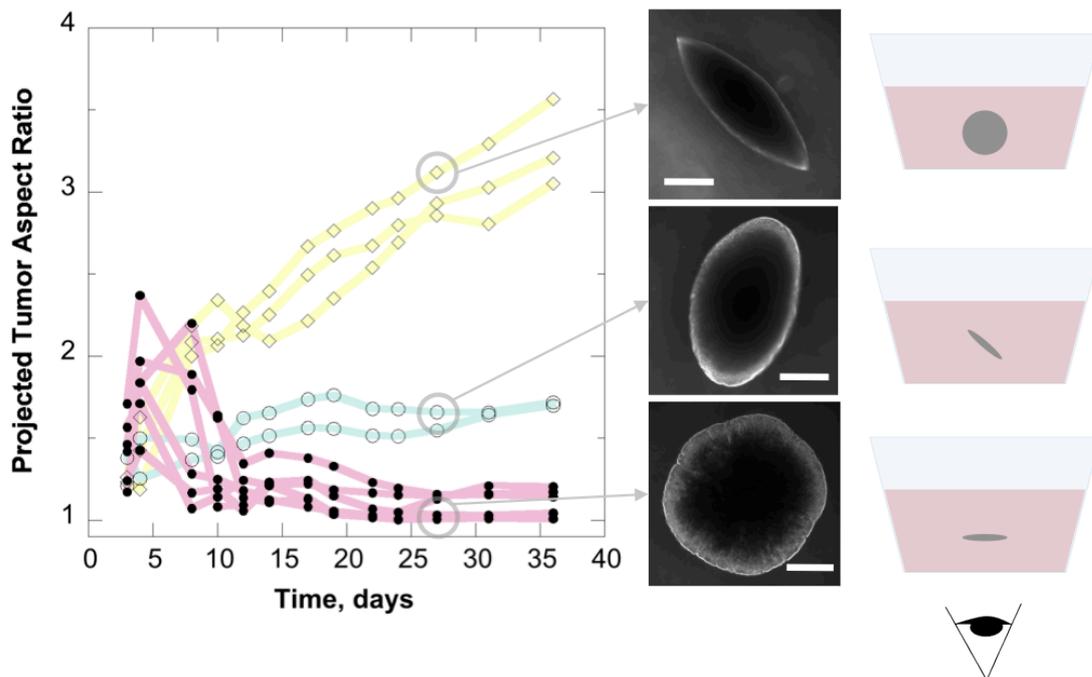

**Figure S4. Individual tumor development.** To determine the time frame over which the oblate ellipsoidal shape developed, individually embedded, pre-produced tumor spheroids were followed as they grew in the agarose gel. In such an experiment, it was determined that within one week after embedment the ellipsoidal shape was detectable and the orientation of the tumor in the gel was fixed. Shown here is the time course development of the projected tumor aspect ratio of 11 different individual tumors (sparsely embedded). Diamond symbols mark the aspect-ratio measurements for tumors whose final orientation projected narrow ellipses, solid circles for tumors whose final orientation projected near-circles, and open circles for tumors whose final orientation projected wider ellipses. Phase-contrast images show the projections from which the aspect ratios were measured for the specific tumors and time points circled in grey on the plot. Scale bars are 500 μm.

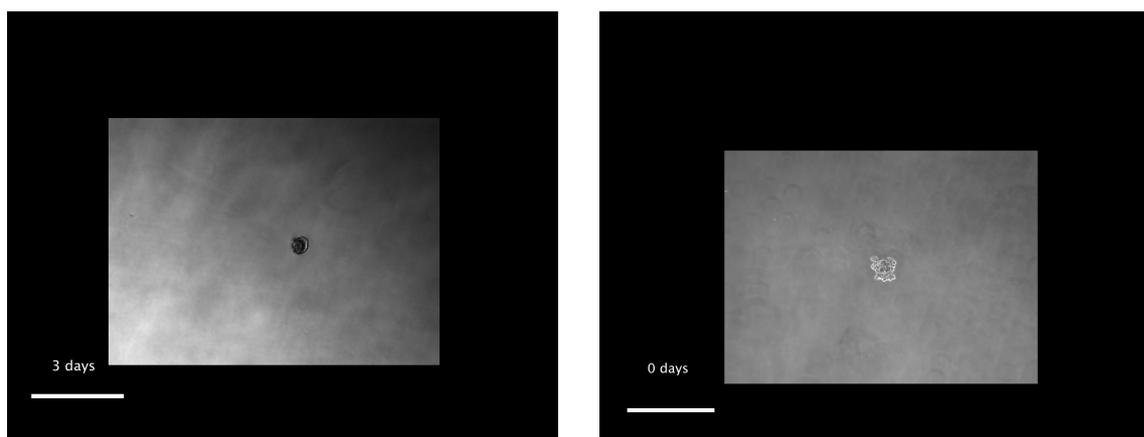

**Movies S1 & S2.** Time lapse progression of individually embedded tumors that developed elliptical and approximately circular cross-sections, respectively. Scale bars are 500 μm.

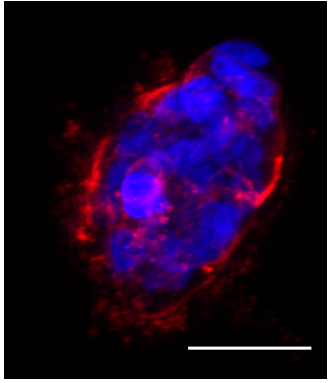

**Figure S5. In the ellipsoidal cross-section of a three-day-old tumor, a necrotic core is not detectable.** It was confirmed that the tumors took on the oblate ellipsoidal shape before the observation of a central necrotic core, ruling out the possibility that this shape formed due to mechanical collapse of a necrosed core. Three-dimensional projection of 13 image slices, taken 2 μm apart in the mid-section of an oblate ellipsoidal tumor. The section is shown here tilted about the vertical axis of the image by 10 degrees. Nuclei (blue) and actin network (red) of LS174T cells embedded in 1% agarose hydrogel. Scale bar is 30 μm.

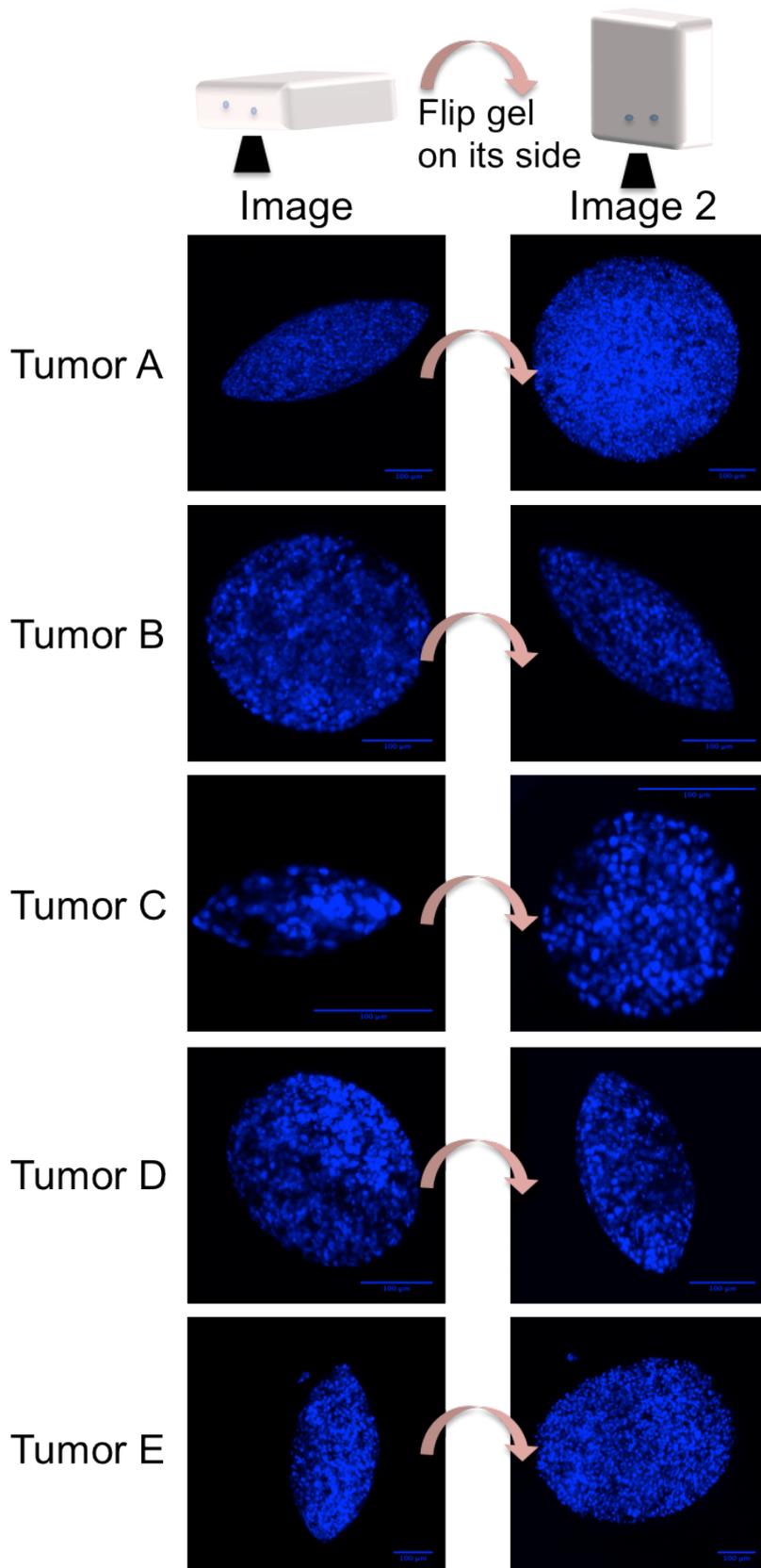

**Figure S6. Perpendicular views of 5 tumors grown in 1% agarose.** Due to scattering regular confocal microscopy is not effective at capturing entire 3D tumor shapes. However, by carefully sectioning the gel in the region of a tumor, two approximately perpendicular views may be imaged of the tumors. Here are some examples. The fluorescent signal is provided by Hoechst-stained nuclei. The scale bars are 100 μm long.

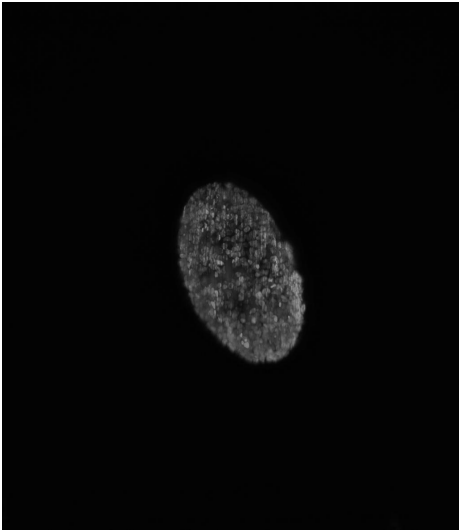
**Movie S3.** 3D rendering of Tumor 1 (Figure 2 in the main paper).

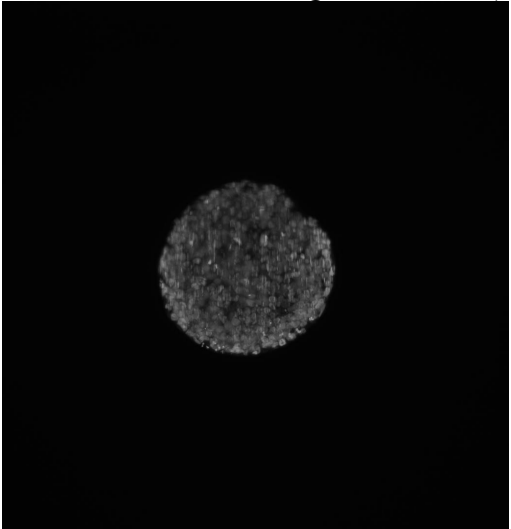
**Movie S4.** 3D rendering of Tumor 2 (Figure 2 in the main paper).